\documentclass[10pt,conference,compsocconf]{IEEEtran}
\IEEEoverridecommandlockouts
\usepackage{listings}
\usepackage{cite}
\usepackage{alltt}
\usepackage{sg}
\usepackage{array}
\usepackage{listings}
\usepackage[caption=false,font=footnotesize]{subfig}
\usepackage{fixltx2e}
\usepackage{amsmath}
\usepackage{booktabs}
\usepackage{colortbl}
\usepackage{float}
\usepackage{graphicx}
\usepackage[hyphens,spaces,obeyspaces]{url}
\usepackage{breakurl} 
\floatstyle{plain}
\newfloat{eqn}{tpb}{lop}
\floatname{eqn}{Equation}

\begin{document}
%
\title{Evaluating Rapid Application Development with Python for Heterogeneous Processor-based FPGAs}

\author{\IEEEauthorblockN{Andrew G. Schmidt, Gabriel Weisz, and Matthew French}
\IEEEauthorblockA{Information Sciences Institute, University of Southern California \\
Email: \{aschmidt, gweisz, mfrench\}@isi.edu}}

\maketitle

\begin{abstract}
As modern FPGAs evolve to include more heterogeneous processing
elements, such as ARM cores, it makes sense to consider these devices
as processors first and FPGA accelerators second.  As such, the
conventional FPGA development environment must also adapt to support
more software-like programming functionality.  While high-level
synthesis tools can help reduce FPGA development time, there still
remains a large expertise gap in order to realize highly performing
implementations.  At a system-level the skill set necessary to
integrate multiple custom IP hardware cores, interconnects, memory
interfaces, and now heterogeneous processing elements is complex.
Rather than drive FPGA development from the hardware up, we consider
the impact of leveraging Python to accelerate application development.
Python offers highly optimized libraries from an incredibly large
developer community, yet is limited to the performance of the hardware
system.  In this work we evaluate the impact of using PYNQ, a Python
development environment for application development on the Xilinx Zynq
devices, the performance implications, and bottlenecks associated with
it.  We compare our results against existing C-based and hand-coded
implementations to better understand if Python can be the glue that
binds together software and hardware developers.

\end{abstract}

\section{Introduction}
\label{sec:intro}

As FPGA devices continue to increase in heterogeneity,
e.g.\ incorporating multi-core ARM processors, the software community
has been taking notice.  Moreover, industry has been shifting its
focus to FPGAs over the past few years, as evidenced by Microsoft's
Catapult project\cite{Catapult}, Intel's acquisition of
Altera\cite{IntelAltera}, and most recently Amazon's inclusion of
FPGAs as part of their Amazon Web Services\cite{AmazonF1}.  While the
FPGA community is keenly aware of the performance and power
efficiencies FPGAs offer developers, there remains a significant
challenge to broaden FPGA usage.  High-level synthesis (HLS) and other
productivity tools are a start, but still require FPGA expertise to
direct the tools to achieve good results.  While HLS has an important
role in FPGA development, incorporating hardware accelerators into an
end-user's application can be a daunting task.  The software community
is use to leveraging efficient libraries, highly tuned for the
hardware in order to obtain the best performance.  What the FPGA
community needs to embrace is a more software-down development flow
rather than hardware-up.  Furthermore, for wider FPGA adoption by the
software community, the tools and languages supported need to go
beyond conventional embedded systems languages.

Over the last several years Python has grown in popularity both in
academia and industry\cite{langList}.  With a wide variety of
libraries and tools available to developers, Python is being used in
everything from scientific computing to image processing and machine
learning, and growing more each day.  Making FPGAs more user friendly
certainly has been an on-going effort for decades and this work does
not claim to solve this problem.  Instead, it looks at how entire
communities have sprung up seemingly overnight around other embedded
platforms, such as Raspberry Pi and Arduino.  The success of these
platforms stems from an inexpensive compute platform, ease of use
programming environment, modularity, and a plethora of interesting and
fun projects readily available to be tried, modified, and refined.

Towards this trend, Xilinx recently released PYNQ (PYthon on zyNQ)
\cite{pynq} as a productivity environment and platform for developers,
combining the use of Python, its tools and libraries with the
capabilities of programmable logic and ARM processors.  High-level
languages are desired in an embedded space where today C/C++
dominates, so long as the performance is not impacted.  This paper
aims to leverage Python for rapid application development on FPGAs and
to understand the performance and development implications of doing
so.  With Python we can quickly develop an application, in this case
Edge Detection, and compare the performance across several different
C/C++, Python, and hardware accelerated implementations.  Our results
are highly encouraging in that not only can using Python reduce
application development time by exploiting a tremendously rich and
diverse set of packages, libraries, and tools, but we are also able to
obtain highly performing implementations when compared to conventional
C/C++ embedded implementations.



\section{Background and Related Work}
\label{sec:back}

With the goal of this paper being to explore how an application
developer might utilize custom hardware kernels with the Xilinx PYNQ
application framework\cite{pynq}, it is important to describe what
PYNQ is and how this work is using Python.  The PYNQ application
development framework is an open source effort designed to allow
application developers to achieve a ``fast start'' in FPGA application
development through use of the Python language and standard
``overlay'' bitstreams that are used to interact with the chip's I/O
devices.  The PYNQ environment comes with a standard overlay that
supports HDMI and Audio inputs and outputs, as well as two 12-pin PMOD
connectors and an Arduino-compatible connector that can
interact with Arduino shields. The default overlay instantiates
several MicroBlaze processor cores to drive the various I/O
interfaces. Existing overlays also provide image filtering
functionality and a soft-logic GPU for experimenting with SIMT-style
programming\cite{fgpu}.  PYNQ also offers an API and extends common
Python libraries and packages to include support for Bitstream
programming, directly access the programmable fabric through
Memory-Mapped I/O (MMIO) and Direct Memory Access (DMA) transactions
without requiring the creation of device drivers and kernel modules.
Our work builds upon these APIs and Overlay concepts to develop
application kernels that can be dynamically connected together to
create processing pipelines.

Several existing projects \cite{pymetal, pyhdl, syspy, myhdl} allow
application developers to describe hardware kernels (and even entire
systems) using low-level python code. This approach is complementary
to our approach, in that these projects could be used to create
hardware kernels that can be incorporated into PYNQ overlays. These
systems utilize a Python syntax to describe hardware in a way that is
functionally equivalent to behavioral HDL, and are not as
sophisticated in terms of the code that they accept as modern C-based
high-level synthesis tools such as Vivado HLS, which could also be
used to generate hardware kernels that are connected together using
our approach.

\section{Design}
\label{sec:design}

There exist a number of approaches and conventions for embedded system
development.  The common approach includes implementing a design in
C/C++, profiling the application to determine the computationally
intensive portions, and migrating those kernels to hardware through
either custom HDL or a high-level synthesis tool.  While C/C++ remain
near the top of the list of programming languages for embedded
systems, Python has consistently been ranked at or near the top of
Lists of Programming languages taught in academia and used in
industry.  As a result, we consider what design and performance
implications are involved when using Python in an FPGA development
environment.  This process in motivated by the release of PYNQ from
Xilinx which aids in the interfacing with custom hardware in the FPGA
fabric and providing a number of useful utilities, such as downloading
bitstreams from within the application.

\begin{figure}
\begin{center}
\includegraphics[width=0.9\linewidth]{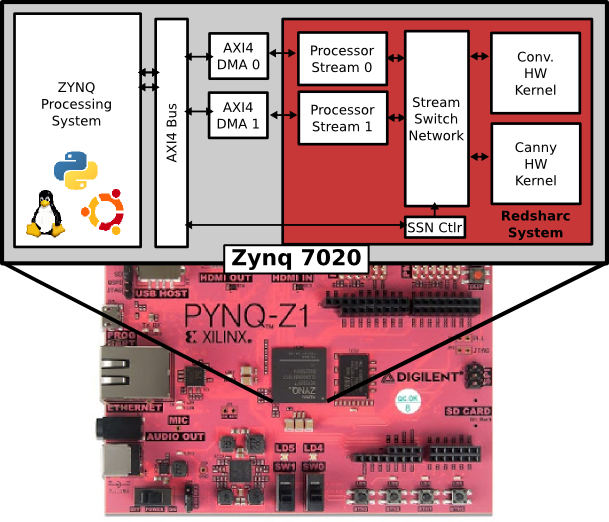}
\caption{Redsharc system for edge detection running on PYNQ}
\label{fig:Redsharc_PYNQ}
\end{center}
\vspace{-0.5cm}
\end{figure}

First we must consider what PYNQ is and is not.  PYNQ does not
currently provide or perform any high-level synthesis or porting of
Python applications directly into the FPGA fabric.  As a result, a
developer still must use create a design using the FPGA fabric.  While
PYNQ does provide an Overlay framework to support interfacing with the
board's IO, any custom logic must be created and integrated by the
developer.  A developer can still use high-level synthesis tools or
the aforementioned Python-to-HDL projects to accomplish this task, but
ultimately the developer must create a bitstream based on the design
they wish to integrate with the Python, seen in
Figure~\ref{fig:Redsharc_PYNQ}.

\begin{figure}[b]
\begin{center}
\includegraphics[width=0.9\linewidth]{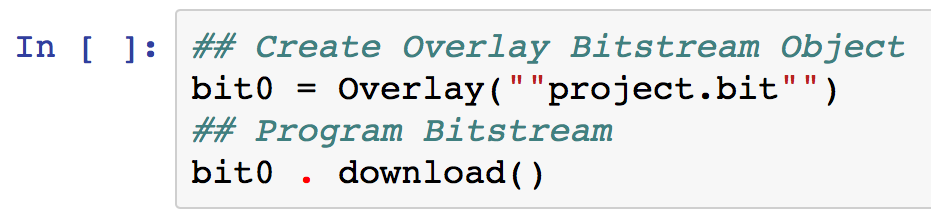}
\caption{PYNQ Programming Bitstream Example}
\label{fig:PYNQprog}
\end{center}
\vspace{-0.5cm}
\end{figure}

What PYNQ does provide is a simplified way of integrating and
interfacing with the hardware once it is designed and the bitstream is
created, for example bitstream programming as shown in
Figure~\ref{fig:PYNQprog}.  Plus, PYNQ exposes the wealth of
additional Python libraries and tools to allow for a much richer
software development environment than conventional C/C++ embedded
systems design.  This includes interactive debuggers, \textit{pdb},
profiling and measurement tools, \textit{cProfile/timeit}, and
libraries and packages like \textit{NumPy}, \textit{SciPy}, and
\textit{matplotlib}.


In this work typical FPGA development is still necessary in that a
Vivado project is created, hardware accelerators are added and the
design is synthesized, implemented, and a bitstream is generated.
PYNQ does not change this process.  For traditional FPGA developers
this is actually a comforting fact, which means existing designs and
tools do not necessarily need to be modified to work with PYNQ.
Existing overlays or hardware/software co-design tools that assemble a
bitstream through Vivado will still work.  While a number of different
hardware/software development environments exist
\cite{hthreads,reconos,leap,coram} this work uses the Redsharc
project \cite{redsharc} due to its focus on streaming-based kernel
development and tight integration with the Xilinx tool-flow.

\begin{figure}[H]
\begin{center}
\includegraphics[width=0.99\linewidth]{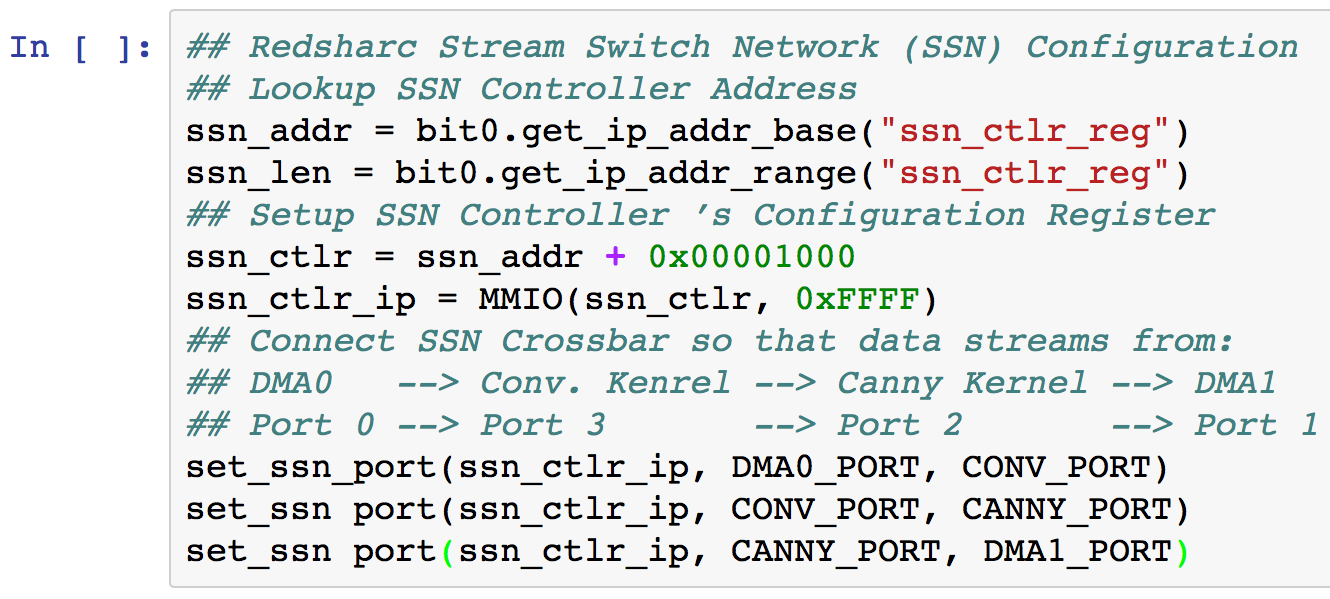}
\caption{Configuring Redsharc with PYNQ}
\label{fig:RedsharcPYNQ}
\end{center}
\vspace{-0.5cm}
\end{figure}

Within Redsharc the hardware kernel development is simplified by
abstracting away the complexities of a full system-on-chip design.
This is accomplished by handling the system assembly, run-time
management, and data transfers, for the designer.  In effect, the
developer is now tasked with creating high performance compute kernels
much like how highly efficient libraries are developed and leveraged
in Python.  Redsharc can then be integrated within the PYNQ
application through simple MMIO functions to configure the
connectivity of the different hardware kernels and DMA cores.  PYNQ
uses the C Foreign Function Interface for Python (CFFI)\cite{cffi}, a
standard Python library, to bind with any existing C shared object
libraries, like the DMA controller.  An example of this setup and
configure is shown in Figure~\ref{fig:Redsharc_PYNQ} and in
Figure~\ref{fig:RedsharcPYNQ}.




\begin{figure}[H]
\begin{center}
\includegraphics[width=0.8\linewidth]{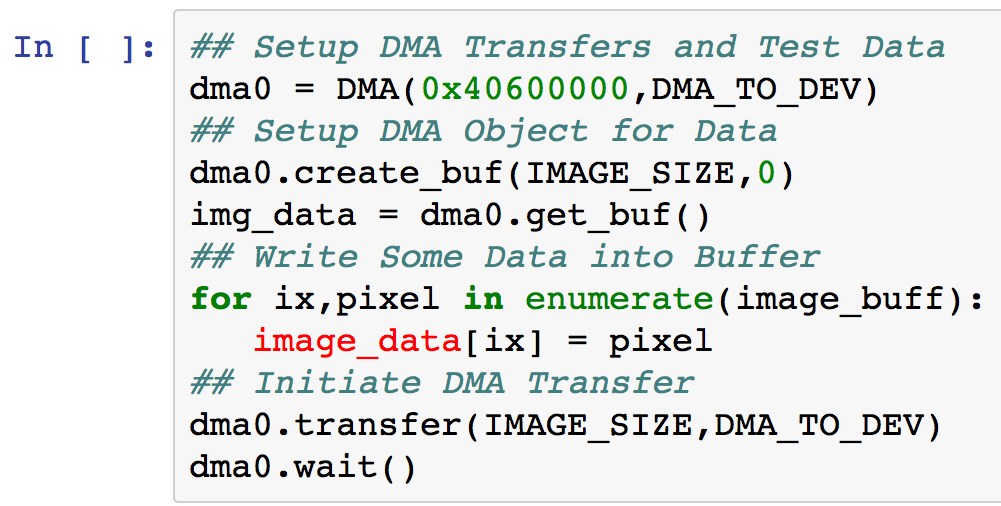}
\caption{DMA example with PYNQ}
\label{fig:PYNQDMA}
\end{center}
\end{figure}

In addition to MMIO, PYNQ provides a convenient and efficient way to
perform DMAs between memory and the programmable fabric.  The DMA
engine is first initialized, then a buffer is created and can be
interfaced in any way the user needs.  Once ready for the transfer,
the user can call a simple transfer for the DMA, all shown in
Figure~\ref{fig:PYNQDMA}.





\section{Evaluation}
\label{sec:results}

To understand and evaluate the performance implications of using
Python and PYNQ for application development we use and compare C,
Python, OpenCV libraries, and custom hardware accelerators.  This
section first describes the different testing configurations of the
experimental setup followed by the analysis and discussion of the
results.

\subsection{Experimental Setup}

For this work we conducted several experiments on the Xilinx PYNQ
platform\cite{pynq}, which includes the Xilinx xc7z020clg400-1 part
and 512 MB of DDR3 memory.  The processor clock is configured for
667~MHz and the fabric and hardware accelerators are configured to run
at 200~MHz.  Each experiment performed Edge Detection on 1024x768
grayscale images, a common step in many image processing
pipelines~\cite{rockster}.  Our motivation for using Edge Detection is
the widely available code and libraries, as well as, being a highly
useful feature of image processing flows.

In total six different software and hardware configurations are used
in this experiment.  The purpose is to evaluate the performance
implications of using C vs. Python in an embedded development
environment with FPGAs for application development.  The hardware for
these experiments include a custom 2D direct convolution kernel for
Gaussian filtering, and a publicly available Canny edge detector core
that performs the gradient calculation and non-max suppression
steps~\cite{opencorescanny}, modified to improved buffering.  The
hardware kernels each use streaming interfaces that can consume and
produce 1 pixel per cycle, using 32-bit integer accumulation during
convolution, and 32-bit integer gradient calculation.  The FPGA is
configured the same for both C and Python-based experiments.

The C versions were written using OpenMP and run on one and two
threads to utilize the two ARM A9 cores on the Zynq 7020 device.  The
OpenCV version utilizes the OpenCV library to perform image
convolution using the \textit{GaussianBlur} function followed by the
\textit{Canny} function.  The hardware accelerated version utilizes a
hand-coded convolution and canny edge detector kernel running at 200
MHz in the FPGA fabric.  The C versions is our baseline and shows what
a number of research papers have already shown, edge detection on
FPGAs can offer performance improvements over software
implementations.

\subsection{Results and Analysis}

The results of running Edge Detection on six different hardware and
software configurations is shown in Table~\ref{tab:results}.  First,
we show the performance gains from traditional C implementations on
one and two cores.  Using OpenMP we are able to nearly achieve linear
speedup from one to two cores.  With OpenCV we are able to leverage
highly optimized software implementations for the kernels and achieve
an impressive 22.91$\times$ speedup over the C reference
implementation.  The hardware accelerated version does slightly
outperform the OpenCV version by streaming the output of the
convolution kernel directly into the canny kernel, without requiring a
memory transaction.  The work involved to achieve these performance
gains did require development effort.  Integrating OpenMP to provide
better scalability across the ARM A9 cores took approximately one day.
The OpenCV implementation was based on reference designs online, but
did require cross-compiling and installing the necessary libraries on
the target platform.  The entire process was performed in
approximately two hours.  Finally, the hardware accelerated version
leveraged an Open Source implementation, but in order to obtain better
throughput a buffering mechanism was added.  The hardware
implementation took approximately one week.  These efforts could have
been improved by using high-level synthesis tools, and as such, is not
meant to be a main takeaway from this work.

\begin{table}
\begin{center}
\caption{Experimental results comparison for edge detection}
\label{tab:results}
\begin{tabular}{lcc}
\hline
\multicolumn{1}{c}{\textbf{Configuration}} & \textbf{Time (s)} & \textbf{Speedup}\\
\hline
\hline
\rowcolor[gray]{0.9} C Version - 1 Thread        & 2.0516     & 1.00$\times$ \\
                     C Version - 2 Threads       & 1.0660     & 1.93$\times$ \\
\rowcolor[gray]{0.9} OpenCV Version - 2 Threads  & 0.0896     & 22.91$\times$ \\
                     HW Accelerated Version      & 0.0765     & 26.80$\times$ \\
\hline
\rowcolor[gray]{0.9} Python OpenCV Version       & 0.1795     & 11.43$\times$ \\
                     PYNQ HW Accelerated Version & 0.0679     & 30.21$\times$ \\
\hline
\end{tabular}
\end{center}
\vspace{-0.5cm}
\end{table}

Instead, we focus on the ability to rapidly develop an application and
obtain results on the target platform.  In the C development
environment this includes compiling and testing on the host,
cross-compiling, testing, and debugging on the target platform, then
integrating with the hardware kernel through device drivers and
possibly other kernel modifications.  The system complexity of
managing kernel, device drivers, root file systems, data in kernel
space vs. user space, in addition to the FPGA development quickly
necessitates a broad skill set or a development team.

Utilizing a platform such as PYNQ, where Python is the main
programmers interface to the hardware, the portability complexity and
need for cross-compiler and device drivers is eliminated.  PYNQ
provides APIs for programming the bitstream, reading and writing data
through MMIO and DMA, significantly reduce the system complexity.  The
profiling and debugging tools built into Python or available through
libraries and package installations enables a developer to quickly
build, test, and refine their application.  

While obtaining performance gains in C and hardware are common place,
we were mostly interested in what the performance and overhead of
using Python and PYNQ.  As software developers embrace Python for ease
of programming, we show that naive ports or implementations can yield
terrible performance.  A straight port of the C version of the Edge
Detector was implemented in Python and resulted in running
334.8$\times$ slower than the C version.  Even though the port took
less than one hour, it is meant to highlight the importance of using
Python's extremely large community of libraries, analysis tools, and
debuggers.  With very little effort, less than 10 minutes, a Python
OpenCV implementation running on the ARM A9 cores, obtaining an
11.43$\times$ speedup over the C version and comical 3,826.94$\times$
speedup over the Python C ported version.

Finally, we wanted to see how a hardware accelerated core combined
with Python would perform.  The speedup is 30.2$\times$ when comparing
with the single threaded C version.  The configuration was even
2.64$\times$ faster when compared against the Python OpenCV version.
Furthermore, when comparing the two hardware versions, C and Python,
it is the Python version that was able to edge out C with a slight
1.12$\times$ improvement.  The differences are largely attributed to
the DMA bandwidth we were achieving, with a slight improvement in the
Python version.  These results are highly encouraging and indicate
that Python in combination with hardware accelerator kernels can match
or even outperform C implementations.



\section{Conclusion and Future Work}
\label{sec:conc}
With FPGAs becoming more heterogeneous, capable, and processor-centric
it is evident a more software-down development environment is needed.
Xilinx recently released PYNQ with the aim to support software
developers using Python to access the FPGA.  The combining of both
Python software and FPGA's performance potential is a significant step
in reaching a broader community of developers, akin to Raspberry Pi
and Ardiuno.  This work studied the performance of common image
processing pipelines in C/C++, Python, and custom hardware
accelerators to better understand the performance and capabilities of
a Python + FPGA development environment.  The results are highly
promising, with the ability to match and exceed performances from C
implementations, up to 30$\times$ speedup.  Moreover, the results show
that while Python has highly efficient libraries available, such as
OpenCV, FPGAs can still offer performance gains to software
developers.

This initial study provides insight into how PYNQ works and how to
interact with the programmable fabric and hardware accelerators
through Python.  The performance results are encouraging and we are
currently evaluating additional application benchmarks in a variety of
scientific computing and machine learning domains.  We are also
evaluating porting the system to the newly released Xilinx Zynq
UltraScale+ FPGA which include four ARM A53 application processors and
two ARM R5 real-time processors.



\bibliographystyle{IEEEtran}
\bibliography{fccm2017}

\end{document}